\documentclass[preprintnumbers,floats,floatfix,amssymb,prd,twocolumn,nofootinbib]{revtex4}
\usepackage{amssymb,amsmath}
\usepackage{epsfig,hyperref}
\usepackage[svgnames]{xcolor}
\usepackage{pgf,tikz}
\usepackage{epstopdf}
\usepackage[british]{babel}
\usepackage{enumerate}
\makeatletter
\DeclareRobustCommand*{\bfseries}{%
  \not@math@alphabet\bfseries\mathbf
  \fontseries\bfdefault\selectfont
  \boldmath
}
\makeatother

\def\be{\begin{equation}}
\def\ee{\end{equation}}
\def\beq{\begin{eqnarray}}
\def\eeq{\end{eqnarray}}
\makeatletter

\newcommand{\arXiv}[2][]{\href{http://arxiv.org/abs/#2}{\texttt{arXiv:#2\@ifempty{#1}{}{ [#1]}}}}
\makeatother

\begin{document}
\title{Spherical vacuum and scalar collapse for the Starobinsky $R^2$ model}

\author{Jun-Qi Guo}%
\email{junqi.guo@tifr.res.in}
\author{Pankaj S. Joshi}%
\email{psj@tifr.res.in}
\affiliation{Department of Astronomy and Astrophysics, Tata Institute of Fundamental Research, Homi Bhabha Road, Mumbai 400005, India}
\date{\today}

\begin{abstract}
  Spherical vacuum and scalar collapse for the Starobinsky $R^2$ model is simulated. Obtained by considering the quantum-gravitational effects, this model would admit some cases of singularity-free cosmological spacetimes. It is found, however, that in vacuum and scalar collapse, when $f'$ or the physical scalar field is strong enough, a black hole including a central singularity can be formed. In addition, near the central singularity, gravity dominates the repulsion from the potential, so that in some circumstances the Ricci scalar is pushed to infinity by gravity. Therefore, the semiclassical effects as included here do not avoid the singularity problem in general relativity. A strong physical scalar field can prevent the Ricci scalar from growing to infinity. Vacuum collapse for the $R\ln{R}$ model is explored, and it is observed that for this model the Ricci scalar can also go to infinity as the central singularity is approached. Therefore, this feature seems universal in vacuum and scalar collapse in $f(R)$ gravity.
\end{abstract}
\maketitle

\section{Introduction\label{sec:introduction}}
Despite the great success of general relativity, many efforts have been made to extend it with the goal of addressing the singularity problem and constructing a quantum theory of gravity. As a spacetime singularity is approached, all physical quantities including the mass-energy density and spacetime curvatures diverge, and the classical theories of gravity break down. Such singularities occur at the origin of the Universe and in gravitational collapse of massive stars when they exhaust their internal fuel (see, e.g., Refs.~\cite{Hawking_1973,Joshi_1993} and references therein for further details). It is widely believed that the singularity problem may be cured through incorporation of nontrivial quantum effects. Attempts have been made in this direction. While we do not have a full quantum theory of gravity as of yet, attempts have been made to take into account the quantum inputs close to the singularity in terms of semiclassical effects, either by quantizing the matter part in the Einstein equations or by quantizing certain limited degrees of freedom of the metric tensor.

The possibility of avoiding the singularity problem by considering a generalized Lagrangian density for gravity in the cosmological context has been investigated in the literature. One of the versions was proposed in Ref.~\cite{Sakharov_1967},
\be
\begin{split}
f(R)=&\Lambda + AR+BR^2+CR^{ik}R_{ik}\\
&+DR^{iklm}R_{iklm}+ER^{iklm}R_{ilkm},
\end{split}
\label{general_f_R}\ee
where $\Lambda$, $A$, $B$, $C$, $D$, and $E$ are constants, and $R$, $R^{ik}$, and $R^{iklm}$ are the Ricci scalar, Ricci tensor, and Riemann tensor, respectively. Considering that not all quadratic invariants in (\ref{general_f_R}) are independent, a simpler form of the Lagrangian density was used in Ref.~\cite{Ruzmaikina_1970},
\be f(R)=\Lambda +AR+BR^2+CR^{ik}R_{ik}.\label{simpler_f_R}\ee
The problem of the nonsingular transition from contraction to expansion for a homogeneous and isotropic cosmological model was investigated. It was found that
the singularity problem remains in the physically acceptable solutions for this model. In Ref.~\cite{Barrow_1983}, considering the model
$f(R)=\Lambda + AR+BR^2$, the conditions for the existence and stability of the de Sitter and Friedman solutions were proven, and the conditions for the existence of cosmological singularities were obtained. The results for cosmological singularities are in agreement with those reported in Ref.~\cite{Ruzmaikina_1970}.

Using the renormalized theory of gravitation, Nariai and Tomita obtained~\cite{Nariai_1971a,Nariai_1971b}
\be f(R)=R+\eta(R^2+\tilde{\alpha}R_{\mu\nu}R^{\mu\nu}),\ee
where $\eta$ and $\tilde\alpha$ are constants. The possibility of singularity avoidance was discussed, and the future de Sitter expansion was found. Taking into account the interaction of quantum free matter fields with a classical gravitational field, Starobinsky obtained an inflationary cosmological model ~\cite{Starobinsky_1980}. An approximate local action for this model can be written as~\cite{Vilenkin_1985,Shore_1980}
\be f(R)=R+\frac{R^2}{6M^2}+\frac{R^2}{R_0}\ln\frac{R}{R_0},\ee
where $M$ and $R_0$ are parameters. For a detailed study of the Starobinsky scenario, see Ref.~\cite{Vilenkin_1985}. For a comparison of the cosmological solutions derived by Nariai and Tomita in Refs.~\cite{Nariai_1971a,Nariai_1971b} and by Starobinsky in Ref.~\cite{Starobinsky_1980}, see Ref.~\cite{Tomita_2016}. A simplified model was studied in Refs.~\cite{Whitt_1984,Wands_1994},
\be f(R)=R+{\alpha}R^2.\label{R_squared_model}\ee
It is a modification of general relativity at high curvature scale and is reduced to general relativity at low curvature scale.
In Refs.~\cite{Whitt_1984,Maeda_1988,Maeda_1989,Barrow_1988,Barrow_1988_2,Barrow_1990,Tsujikawa1}, this model was shown to be equivalent to general relativity with a massive scalar field. It continues to attract considerable interest in that it could drive the inflation in the early Universe~\cite{Starobinsky_1980,Vilenkin_1985,Maeda_1988,Barrow_1988} and is consistent with the Planck 2015 data on the cosmic microwave background~\cite{Planck_1,Planck_2}. The parameter $\alpha$ can be rewritten as $\alpha=1/(6M^{2}m^{2}_{\tiny \mbox{pl}})$, where $m_{\tiny \mbox{pl}}$ is the Planck mass. The normalization of the cosmic microwave background anisotropies determines $M\approx10^{-5}$~\cite{Ade_5082,Kehagias_1312}.

Going beyond cosmological studies of this model, it is instructive to understand the model's dynamics in terms of gravitational collapse towards black hole formation. In Ref.~\cite{Frolov_2008}, it was pointed out that for dark energy $f(R)$ gravity, in cosmology and static compact stars, a perturbation from the matter field may push the scalar degree of freedom $f'({\equiv}df/dR)$ to $1$. As a result, the Ricci scalar $R$ will become singular. (There is a debate on this subject in Ref.~\cite{Jaime:2010kn}.) In Refs.~\cite{Nojiri_0804,Bamba_0807,Capozziello_0903,Appleby_0909,Bamba_1012,Bamba_1101}, it was argued that this singularity problem can be avoided by adding an $R^2$ term to the original dark energy $f(R)$ function. In this new combined model, a singular $R$ is pushed to regions where $f'$ is also singular. Indeed, when viewed in the Jordan frame, the potential for the combined model is very steep at high curvature scale. Therefore, it is possible that in the late Universe and in static compact stars, the Ricci scalar may remain finite under perturbations. In fact, one can ask in a more challenging circumstance---gravitational collapse towards black hole formation---whether $R$ can go to infinity as the central singularity is approached. Since the combined model is reduced to the $R^2$ model at high curvature scale, for simplicity, we only consider the $R^2$ model in this paper. In Ref.~\cite{Hwang:2011kg}, neutral and charged scalar collapse for the $R^2$ model was simulated. In the neutral collapse case, a black hole including a spacelike central singularity was formed. However, the question of whether $R$ can go to infinity was not investigated. (While we do generally know that inflationary cosmological models have a singularity in the past~\cite{Borde:2001nh}, we know this in terms of the existence of an incomplete geodesic only.)

As a key topic in gravitational physics, gravitational collapse studies have a long history. The Oppenheimer-Snyder solution provides an analytic description of spherical dust collapse into a Schwarzschild black hole~\cite{Oppenheimer}. A spherically symmetric inhomogeneous universe filled with dust matter was explored in Refs.~\cite{Lemaitre,Tolman,Bondi}. The inhomogeneous gravitational collapse of Tolman-Bondi dust clouds was generally investigated in Ref.~\cite{Joshi_1993_PRD}. Simulations of spherical fluid/scalar collapse in general relativity or scalar-tensor theories of gravity (including the Brans-Dicke theory) or AdS spacetimes were implemented in Refs.~\cite{Matsuda,Choptuik,Shibata,Scheel_1,Scheel_2,Harada:1996wt,Hod,Novak_1997,Novak_1999,Hwang_2010,Buchel}. Stability analysis in spherical/cylindrical collapse of an anisotropic charged/neutral fluid in $f(R)=R+{\epsilon}R^n$ gravity was performed in Refs.~\cite{Sharif_2013,Kausar_2014,Sharif_2014}. The constraints from the extra matching conditions in $f(R)$ gravity on collapse of massive stars were studied in Ref.~\cite{Goswami_2014}. Spherical scalar collapse in dark energy $f(R)$ gravity towards black hole formation was simulated in Ref.~\cite{Guo_1312}. Dark matter halo formation in $f(R)$ gravity and Galileon gravity was discussed in Refs.~\cite{Kopp} and \cite{Barreira}, respectively. Although many studies have been done on scalar/fluid collapse in modified gravity, to a large extent, a simpler case, vacuum collapse (where physical fields are absent) in modified gravity, has been skipped.

In this paper, we simulate spherical vacuum and scalar collapse for the $R^2$ model. Interestingly, besides scalar collapse~\cite{Hwang:2011kg}, in vacuum collapse, when the scalar degree of freedom $f'$ is strong enough, a black hole including a central singularity can be formed. In addition, we notice that the steepness of the potential at high curvature scale in the Jordan frame is somewhat misleading. After having transformed the $R^2$ model into the Einstein frame, the potential becomes very flat at high curvature scale. Consequently, in addition to a spacelike central singularity formation, under certain initial conditions, as the central singularity is approached, the Ricci scalar can be pushed to infinity easily. So the classical singularity problem remains in collapse for the $R^2$ model. We find that in scalar collapse, when the physical scalar field is strong enough, although a black hole including a central singularity can still be formed, the physical scalar field may stop the Ricci scalar from growing to infinity. Vacuum collapse for the $R\ln{R}$ model~\cite{Frolov_1101,Guo_1305} is also explored. Although the potential for this model in both the Jordan and Einstein frames is very steep at high curvature scale, in certain circumstances, the Ricci scalar can still be pushed to infinity as the central singularity is approached. Therefore, such a feature seems universal in vacuum and scalar collapse for $f(R)$ gravity.

The paper is organized as follows. In Sec.~\ref{sec:framework}, we develop the framework. Vacuum and scalar collapse for the $R^2$ model is explored in Secs.~\ref{sec:vacuum_collapse} and \ref{sec:scalar_collapse}, respectively. In Sec.~\ref{sec:collapse_RlnR}, we consider vacuum collapse for the $R\ln{R}$ model. Results are summarized in Sec.~\ref{sec:summary}. We set $G=c=\hbar=1$. Then the Planck mass, length, and time are $m_{\tiny \mbox{pl}}=\sqrt{\hbar c/G}=1$, $l_{\tiny \mbox{pl}}=\sqrt{\hbar G/c^3}=1$, and $t_{\tiny \mbox{pl}}=\sqrt{\hbar G/c^5}=1$, respectively. We use $l_{\tiny \mbox{pl}}$ and
$t_{\tiny \mbox{pl}}$ as the units of length and time, respectively.

\section{Framework\label{sec:framework}}
\begin{figure}
  \epsfig{file=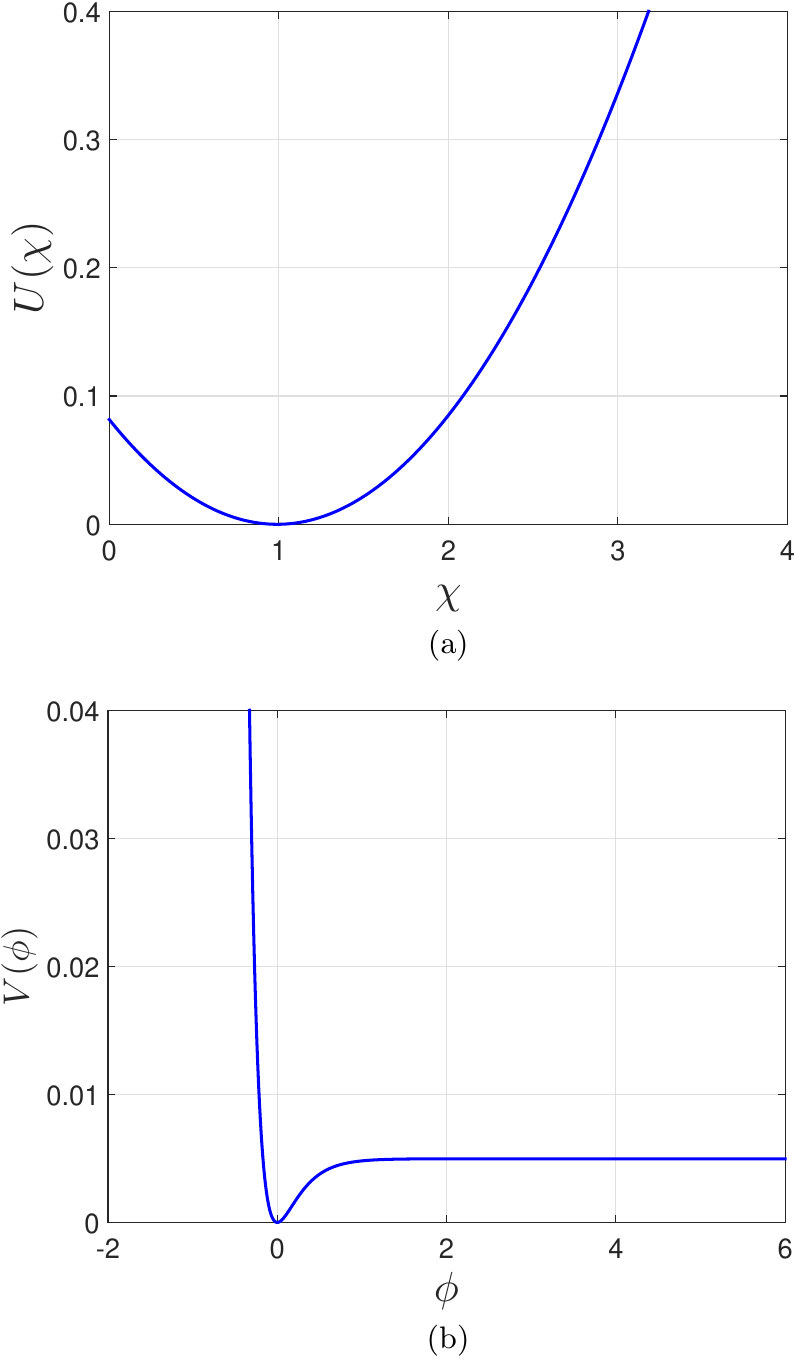,width=6.4cm}
  \caption{Potential for the Starobinsky $R^2$ model~(\ref{R_squared_model}), $f(R)=R+{\alpha}R^2$, with $\alpha=1$. (a) Potential in the Jordan frame~(\ref{U_prime_JF}). (b) Potential in the Einstein frame~(\ref{potential_inflation_model}).}
  \label{fig:potential}
\end{figure}

We start with the action for $f(R)$ gravity
\be S=\int d^{4}x \sqrt{-g}\left[\frac{f(R)}{16\pi G}+\mathcal{L}_{\psi}\right], \label{f_R_action}\ee
where $\mathcal{L}_{\psi}$ is the Lagrangian density for a massless physical scalar field $\psi$, $\mathcal{L}_{\psi}=-(1/2)g^{\alpha\beta}\psi_{,\alpha}\psi_{,\beta}$. The energy-momentum tensor for this scalar field is
\be T^{(\psi)}_{\mu\nu}=\psi_{,\mu}\psi_{,\nu}-\frac{1}{2}g_{\mu\nu}g^{\alpha\beta}\psi_{,\alpha}\psi_{,\beta}.\label{energy_tensor_psi}\ee
From Eq.~(\ref{f_R_action}), the field equation for $f(R)$ gravity can be obtained as
\be f'R_{\mu\nu}-\frac{1}{2}f g_{\mu\nu} +\left(g_{\mu\nu}\Box-\nabla_{\mu}\nabla_{\nu}\right) f'
= 8\pi T^{(\psi)}_{\mu\nu},\label{gravi_eq_fR} \ee
where $\Box\equiv\nabla_{\alpha}\nabla^{\alpha}$. The trace of Eq.~(\ref{gravi_eq_fR}) describes the dynamics for $\chi({\equiv}f')$,
\be \Box\chi-U'(\chi)-\frac{8\pi}{3}T^{(\psi)}=0,\label{trace_eq}\ee
where a potential $U(\chi)$ is defined as
\be U'(\chi)\equiv\frac{dU}{d\chi}=\frac{2f-\chi R}{3},\label{U_prime_JF}\ee
and $T^{(\psi)}$ is the trace of $T^{(\psi)}_{\mu\nu}$. The potential $U(\chi)$ for the $R^2$ model is plotted in Fig.~\ref{fig:potential}(a). Figure~\ref{fig:potential}(a) shows that the potential is finite on the left side, while it goes to $+\infty$ on the right side. It is tempting to think that in collapse, under certain initial conditions, $f'$ may approach zero as it moves to the left. However, although it can rise high in the right direction at a particular moment, it may eventually come back down due to the force from $U'(f')$.

For computational convenience, we transform $f(R)$ gravity from the Jordan frame into the Einstein frame. Defining $\kappa\phi\equiv\sqrt{3/2}\ln\chi$, one obtains the action of $f(R)$ gravity in the Einstein frame~\cite{Whitt_1984,Maeda_1988,Maeda_1989,Barrow_1988,Barrow_1988_2,Barrow_1990,Tsujikawa1},
\begin{eqnarray}
S_E & = & \int d^4 x\sqrt{-\tilde g}\left[\frac{1}{2\kappa^2} \tilde R -\frac{1}{2} \tilde g^{\mu\nu}\partial_\mu\phi\partial_\nu\phi-V(\phi)\right]\nonumber\\
 && + \int d^4 x\mathcal{L}_M\left(\frac{\tilde{g}_{\mu\nu}}{{\chi}},\psi\right),
\end{eqnarray}
where $\kappa=\sqrt{8\pi G}$, $\tilde g_{\mu\nu}=\chi\cdot g_{\mu\nu}$, $V(\phi)\equiv(\chi R-f)/(2\kappa^{2}\chi^2)$, and a tilde denotes that the quantities are in the Einstein frame. The potential $V(\phi)$ for the $R^2$ model can be written as~\cite{Maeda_1988,Barrow_1988,Barrow_1988_2}
\be V(\phi)=\frac{1}{8\kappa^{2}\alpha}\left(1-e^{-\sqrt{2/3}\kappa\phi}\right)^2,\label{potential_inflation_model}\ee
and is plotted in Fig.~\ref{fig:potential}(b). In Ref.~\cite{Maeda_1988}, it was pointed out that because of the flat plateau of $V(\phi)$ at large $\phi$ and the potential minimum at $\phi=0$ inflation is a natural phenomenon in the $R^2$ model. Because of the steepness on the left side ($\phi<0$) and the flatness on the right side ($\phi>0$) for $V(\phi)$, it is tempting to think that in collapse, under certain initial conditions, $\phi$ may go to $+\infty$ in the right direction, while it may not approach $-\infty$ in the left direction.

\begin{figure*}
  \epsfig{file=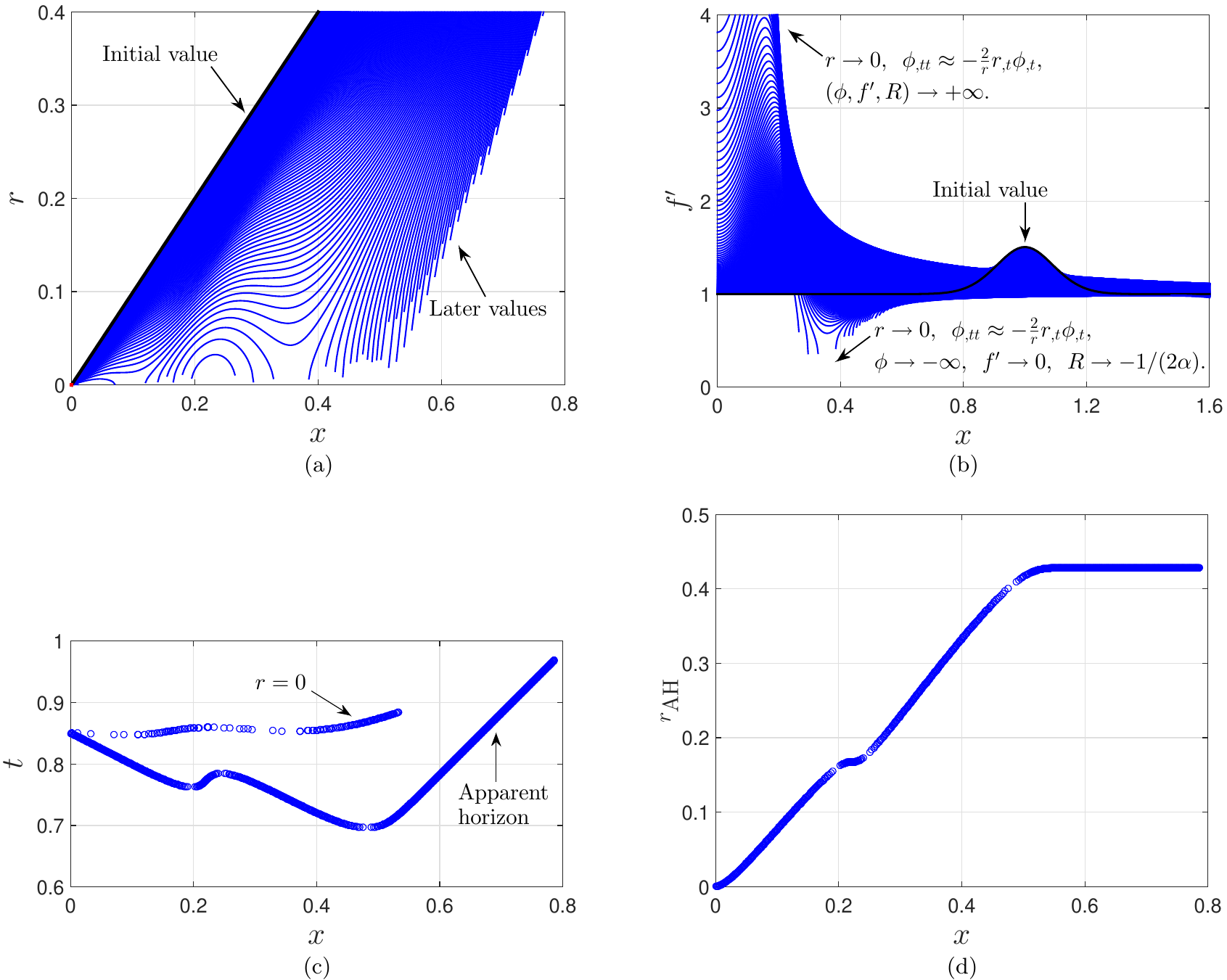, width=0.74\textwidth}
  \caption{Evolutions for vacuum collapse for the Starobinsky $R^2$ model. (a) and (b): Evolutions of $r$ and $f'$, respectively. The time interval between two consecutive slices is $4{\Delta}t=0.002$. (c) and (d): Apparent horizon and singularity curve of the formed black hole. As shown in (a), in later evolutions, the central singularity is approached $(r\rightarrow 0)$. As a result, the equation of motion for $\phi$~(\ref{equation_phi}) is reduced to $\phi_{,tt}\approx-2r_{,t}\phi_{,t}/r$, which describes a positive feedback system since $-2r_{,t}/r$ is positive. Consequently, as shown in (b), around
  $x{\approx} 0.35$ where $\phi_{,t}$ is negative, $\phi$, $f'$, and $R$ are pushed to $-\infty$, $0$, and $-1/(2\alpha)$, respectively; while around $x {\approx} 0.1$ where $\phi_{,t}$ is positive, $\phi$, $f'$, and $R$ are all pushed to $+\infty$. Therefore, the $R^2$ term in Eq.~(\ref{R_squared_model}) does not prevent the singularity formation. The discontinuities of $f'$ in (b) at $x\approx 0.35$ in later evolutions come from the fact that the spacelike singularity $(r=0)$ is met in simulations [also see (a) and (c)].}
  \label{fig:evolutions}
\end{figure*}

\begin{figure*}
  \epsfig{file=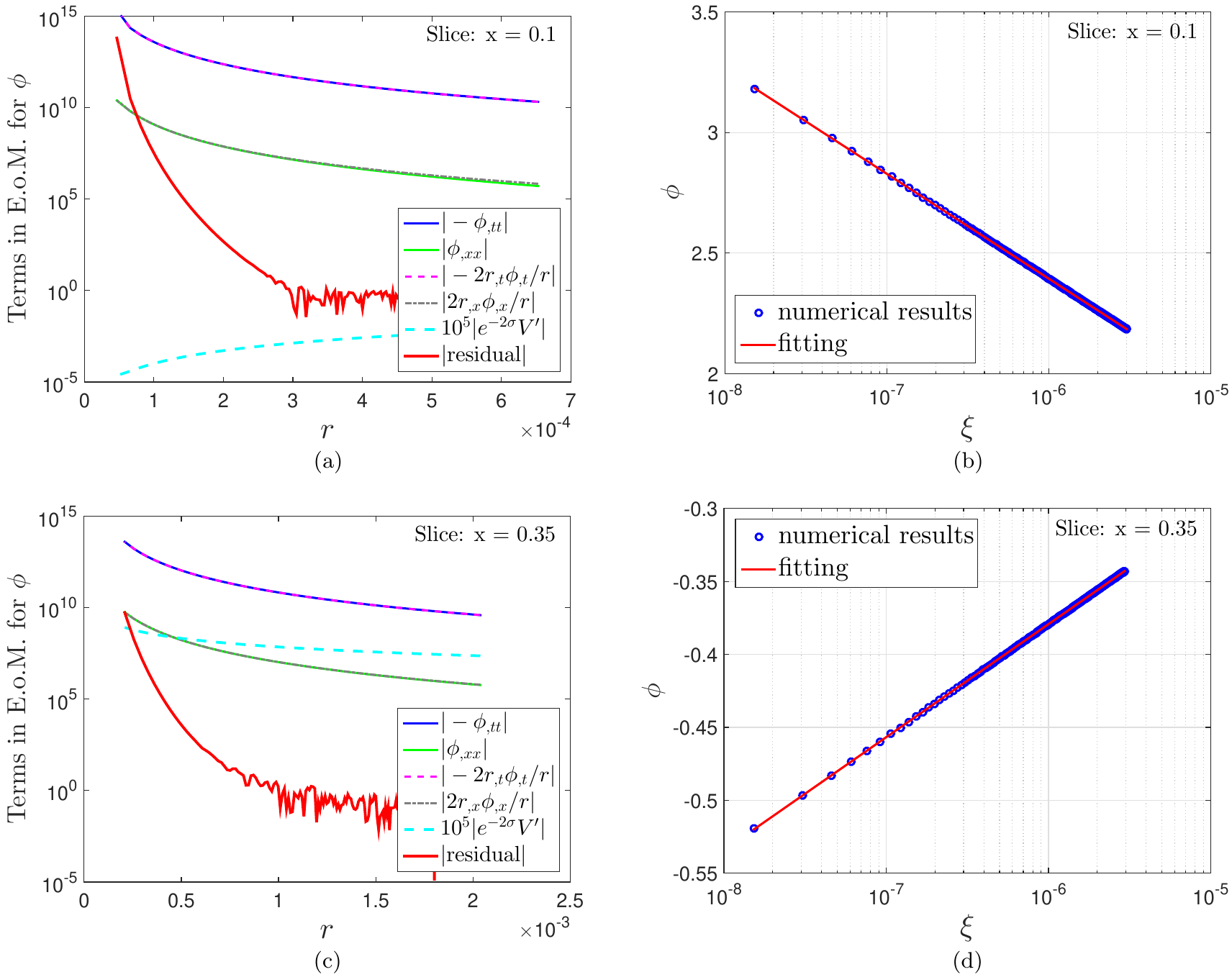, width=0.75\textwidth}
  \caption{(color online). Dynamics for $\phi$ in the vicinity of the singularity curve for vacuum collapse. Here, $\xi=t_0-t$ and $t_0$ is the coordinate time on the singularity curve.
  (a)-(b) and (c)-(d) are for the slices of $x=0.1$ and $x=0.35$, respectively.
  (a) and (c) show that near the singularity there is $\phi_{,tt}\approx-2r_{,t}\phi_{,t}/r$, which is a positive feedback system of $\phi_{,tt}$ and $\phi_{,t}$. Then as shown in (d), on the slice of $x=0.35$, $\phi_{,t}$ is negative, and $\phi$ is pushed to $-\infty$ as the central singularity is approached. However, as shown in (b), on the slice of $x=0.1$, $\phi_{,t}$ is positive, and $\phi$ is pushed to $+\infty$ in later evolutions.
  (b) Here, $\phi=a\ln\xi+b$, $a=-0.18883\pm0.00001$, $b=-0.2152\pm0.0001$.
  (d) Here, $\phi=a\ln\xi+b$, $a=0.033650\pm0.000004$, $b=0.08561\pm0.00006$.}
  \label{fig:dynamics_singularity}
\end{figure*}

The numerical formalism for spherical scalar collapse in $f(R)$ gravity that we use in this paper is essentially the same as the one described in detail in Ref.~\cite{Guo_1312}. Accordingly, we only briefly list the formalism here. We work in the double-null coordinates,
\be
\begin{split}
ds^{2} &= -4e^{-2\sigma}dudv+r^2d\Omega^2\\
&= e^{-2\sigma}(-dt^2+dx^2)+r^2d\Omega^2,
\end{split}
\label{double_null_metric_dtdx}
\ee
where $\sigma$ and $r$ are functions of the coordinates $t$ and $x$, $u=(t-x)/2$, and $v=(t+x)/2$. Then the dynamical equations for $r$, $\sigma$, $\phi$, and $\psi$ can be, respectively, written as follows:
\be r(-r_{,tt}+r_{,xx})-r_{,t}^2+r_{,x}^2 = e^{-2\sigma}(1-8{\pi}r^{2}V),\label{equation_r}\ee
\be
\begin{split}
&-\sigma_{,tt}+\sigma_{,xx}+\frac{r_{,tt}-r_{,xx}}{r}\\
&\quad +4\pi\left(\phi_{,t}^2-\phi_{,x}^2 + \frac{\psi_{,t}^2-\psi_{,x}^2}{\chi}-2e^{-2\sigma}V\right)=0,
\end{split}
\label{equation_sigma}
\ee%
\be
\begin{split}
&-\phi_{,tt}+\phi_{,xx}+\frac{2}{r}(-r_{,t}\phi_{,t}+r_{,x}\phi_{,x})\\
&\quad =e^{-2\sigma}\left[V'+\frac{1}{\sqrt{6}}\kappa\tilde{T}^{(\psi)}\right],
\end{split}
\label{equation_phi}
\ee
\be
\begin{split}
&-\psi_{,tt}+\psi_{,xx}+\frac{2}{r}(-r_{,t}\psi_{,t}+r_{,x}\psi_{,x})\\
&\quad =\sqrt{\frac{2}{3}}\kappa(-\phi_{,t}\psi_{,t}+\phi_{,x}\psi_{,x}),
\end{split}
\label{equation_psi}
\ee
where $r_{,t}\equiv dr/dt$ and other quantities are defined analogously, $V'{\equiv}dV/d\phi$, and $\tilde{T}^{(\psi)}=e^{2\sigma}(\psi_{,t}^2-\psi_{,x}^2)/\chi$. The $\{uu\}$ and $\{vv\}$ components of the Einstein equations work as the constraint equations,
\be r_{,uu}+2\sigma_{,u}r_{,u}+4\pi r\left(\phi_{,u}^2+\frac{\psi_{,u}^2}{\chi}\right)=0, \label{constraint_eq_uu}\ee
\be r_{,vv}+2\sigma_{,v}r_{,v}+4\pi r\left(\phi_{,v}^2+\frac{\psi_{,v}^2}{\chi}\right)=0. \label{constraint_eq_vv}\ee

We set $r_{,tt}=r_{,t}=\sigma_{,t}=\phi_{,t}=\psi_{,t}=0$ at $t=0$. We then define the local Misner-Sharp mass $m$~\cite{Misner} and a new variable $g$, respectively, as
\be g^{\mu\nu}r_{,\mu}r_{,\nu}=e^{2\sigma}(-r_{,t}^2+r_{,x}^2){\equiv}1-\frac{2m}{r},\ee
\be g{\equiv}-2\sigma-\ln(-r_{,u}).\label{g_definition}\ee
Then, the equations for $r$, $m$, and $g$ at $t=0$ can be obtained as below~\cite{Guo_1312,Frolov_2004}:
\begin{align}
r_{,x}&=\left(1-\frac{2m}{r}\right)e^g,\label{r_ic_collapse}\\
%\nonumber\\
m_{,r}&=4{\pi}r^2\left[V+\frac{1}{2}\left(1-\frac{2m}{r}\right)\left(\phi_{,r}^2 + \frac{\psi_{,r}^2}{\chi}\right)\right],\label{m_ic_collapse}\\
%\nonumber\\
g_{,r}&=4{\pi}r\left(\phi_{,r}^2 + \frac{\psi_{,r}^2}{\chi}\right).\label{g_ic_collapse}
\end{align}
We set $r=m=g=0$ at the origin $(x=0,t=0)$. Then the values of $r$, $m$, and $g$ on the initial slice of $t=0$ can be computed by integrating Eqs.~(\ref{r_ic_collapse})-(\ref{g_ic_collapse}) from $x=0$ to the outer boundary via the fourth-order Runge-Kutta method. The values of $r$, $\sigma$, $\phi$, and $\psi$ at $t={\Delta}t$ can be obtained with a second-order Taylor series expansion using Eqs.~(\ref{equation_r})-(\ref{equation_psi}). The value of $g$ at $t={\Delta}t$ can be obtained using the definition for $g$~(\ref{g_definition}). At the inner boundary $x=0$, $r$ is always set to zero. For regularity, we enforce $\phi_{,x}=\psi_{,x}=0$ at $x=0$. The value of $g$ at $x=0$ is obtained via extrapolation. We set up the outer boundary conditions via extrapolation. In fact, as long as the outer boundary is far enough from the center compared to the time scale for the collapse, the outer boundary conditions will not affect the dynamics near the center that we are interested in. The finite difference method and the leapfrog integration scheme are implemented. The numerical code is second-order convergent.

\section{Vacuum collapse for the $R^2$ model\label{sec:vacuum_collapse}}
We first consider vacuum collapse, which can be realized by setting $\psi(x,t)\equiv0$. The initial value for $\phi$ is defined as $\phi(x,t)|_{t=0}=a\cdot\exp\left[-(x-x_1)^2/b\right]$ with $a=0.1$, $b=0.02$, and $x_1=1$. The parameter $\alpha$ in Eq.~(\ref{R_squared_model}) is set to $1$. The range for the spatial coordinate is $x\in[0;\:2]$. The temporal and spatial grid spacings are ${\Delta}t={\Delta}x=5\times10^{-4}$.

We run the simulations. Despite the repulsion from the potential, a black hole is formed. The radius of the apparent horizon of the formed black hole is $r_{\scriptsize{\mbox{AH}}}\approx0.425$ [see Fig.~\ref{fig:evolutions}(d)]. We plot the evolutions of $r$ and $f'$ in Figs.~\ref{fig:evolutions}(a) and \ref{fig:evolutions}(b), respectively. The results for $\sigma$ are similar to those in the collapse for dark energy $f(R)$ gravity plotted in Ref.~\cite{Guo_1312} and are skipped here. The apparent horizon and the singularity curve $r(x,t)=0$ are shown in Figs.~\ref{fig:evolutions}(c) and \ref{fig:evolutions}(d).

The asymptotic dynamics in the vicinity of the central singularity is explored with a mesh refinement as described in Ref.~\cite{Guo_1312}. The slices of $x=0.1$ and $x=0.35$ are chosen as samples. Near the singularity, for $r$, $\sigma$, and $\phi$, the ratios between the spatial and temporal derivatives take similar values and are determined by the slope of the singularity curve $r=0$. This feature and the numerical results enable us to reduce the Einstein equations and the equation of motion for $\phi$~(\ref{equation_r})-(\ref{equation_phi}) as follows~\cite{Guo_1312}:
\begin{align}
rr_{,tt}&\approx-r_{,t}^2,\label{equation_r_asymptotic}\\
%\nonumber\\
\sigma_{,tt}&\approx\frac{r_{,tt}}{r}+4\pi\phi_{,t}^2,\label{equation_sigma_asymptotic}\\
%\nonumber\\
\phi_{,tt}&\approx-\frac{2}{r}r_{,t}\phi_{,t}.\label{equation_phi_asymptotic}
\end{align}
As a representative, the case for $\phi$~(\ref{equation_phi_asymptotic}) is plotted in Figs.~\ref{fig:dynamics_singularity}(a) and~\ref{fig:dynamics_singularity}(c). The solutions to Eqs.~(\ref{equation_r_asymptotic})-(\ref{equation_phi_asymptotic}) are~\cite{Guo_1312}
\begin{align}
r&\approx A\xi^{\beta},\label{r_asymptotic}\\
%\nonumber\\
\sigma&\approx B\ln\xi+\sigma_0,\label{sigma_asymptotic}\\
%\nonumber\\
\phi&\approx C\ln\xi,\label{phi_asymptotic}
\end{align}
with $\beta\approx1/2$ and $B\approx\beta(1-\beta)-4{\pi}C^2$. Here, $\xi=t_0-t$ and $t_0$ is the coordinate time on the singularity curve. The solutions~(\ref{r_asymptotic})-(\ref{phi_asymptotic}) are also confirmed by the numerical results. As a representative, the results for $\phi$ are plotted in Figs.~\ref{fig:dynamics_singularity}(b) and \ref{fig:dynamics_singularity}(d).

The apparent horizon is formed, $r$ can reach zero, and the dynamics near $r=0$ is similar to that near the central singularity of a Schwarzschild black hole. Then the center is a singularity. The numerical results show that, near the central singularity, the potential terms $8{\pi}r^{2}e^{-2\sigma}V$ in Eq.~(\ref{equation_r}) and $e^{-2\sigma}V'$ in Eq.~(\ref{equation_phi}) [see Figs.~\ref{fig:dynamics_singularity}(a) and~\ref{fig:dynamics_singularity}(c)] are negligible compared to the gravitational terms. In other words, when the matter field (which means $f'$ here) is strong enough, the repulsion from the potential cannot prevent the singularity formation.

Near the central singularity, $r_{,t}$ is negative. Accordingly, Eq.~(\ref{equation_phi_asymptotic}) describes a positive feedback system of $\phi_{,tt}$ and $\phi_{,t}$: $|\phi_{,t}|$ produces more of $|\phi_{,tt}|$, and in turn $|\phi_{,tt}|$ produces more of $|\phi_{,t}|$. As shown in Fig.~\ref{fig:dynamics_singularity}(d), on the slice of $x=0.35$, $\phi_{,t}$ is negative. Then as the central singularity is approached in later evolutions, $\phi$, $f'$, and $R$ are pushed to $-\infty$, $0$, and $-1/(2\alpha)$, respectively. These results are similar to those in spherical collapse in dark energy $f(R)$ gravity~\cite{Guo_1312}. However, as shown in Fig.~\ref{fig:dynamics_singularity}(b), on the slice of $x=0.1$, given the configuration that we set, $\phi_{,t}$ obtains a positive value at some moment. Then, in later evolutions, $\phi$, $f'$, and $R$ are all pushed to $+\infty$ as the central singularity is approached. Thus, the repulsion from the potential does not prevent $R$ from becoming singular.

\begin{figure}
  \epsfig{file=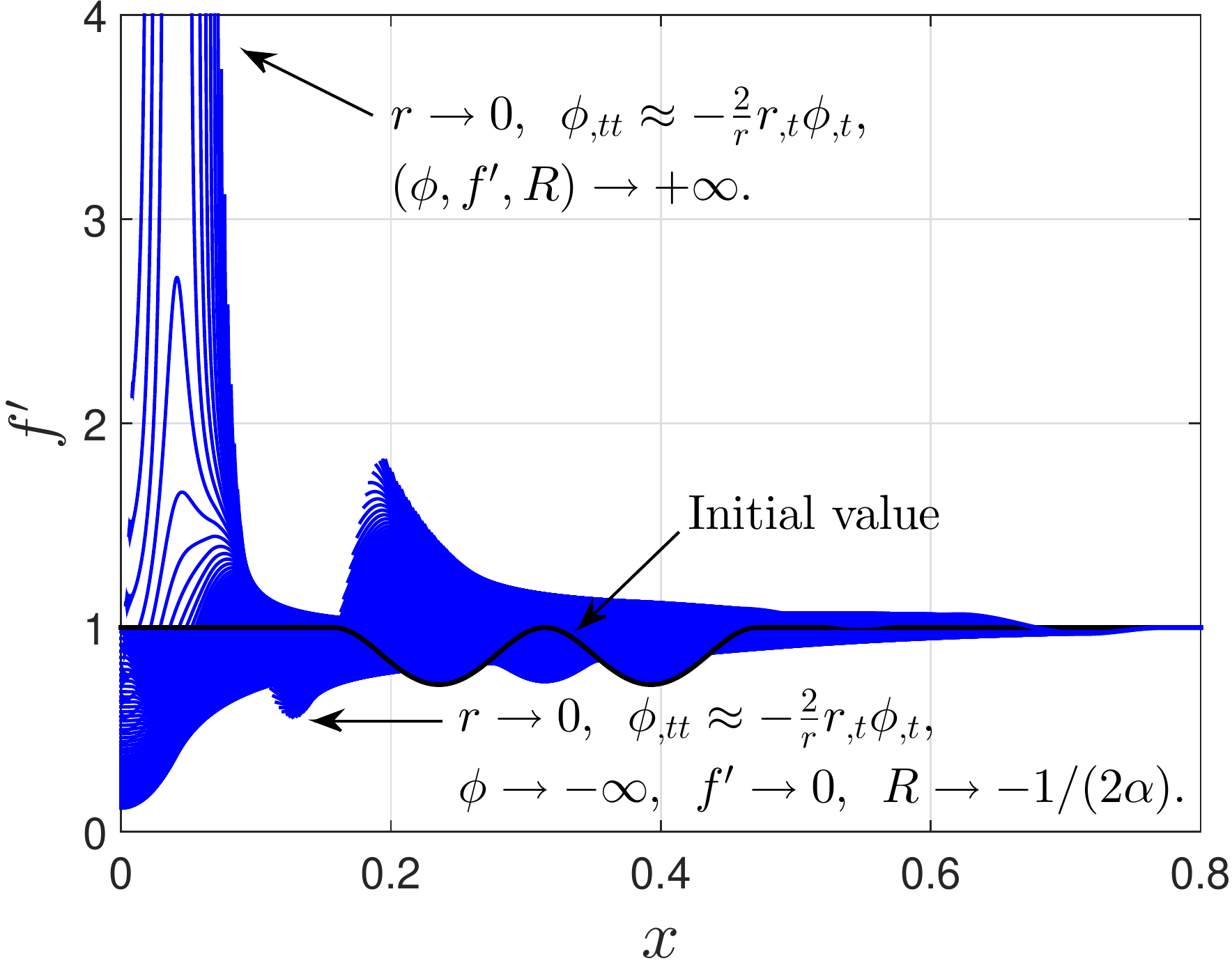,width=6.2cm}
  \caption{Evolutions of $f'$ for vacuum collapse for the Starobinsky $R^2$ model with $\alpha=0.01$. The time interval between two consecutive slices is $2{\Delta}t=4\times10^{-4}$.}
  \label{fig:evolutions_small_alpha}
\end{figure}

\begin{figure*}
  \epsfig{file=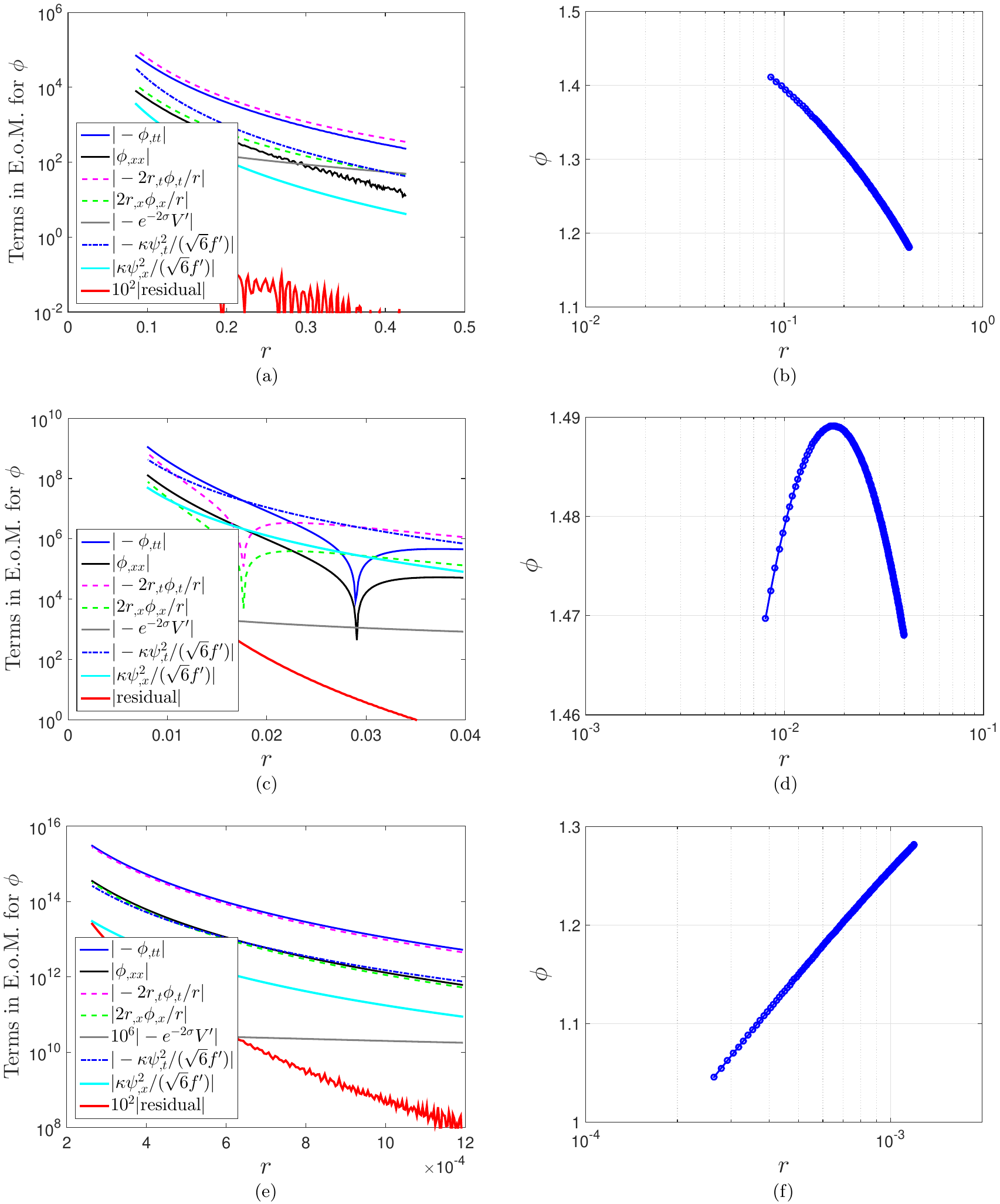, width=0.75\textwidth}
  \caption{(color online). Scalar collapse for the $R^2$ model. The results are for the slice $x=2.8$. (a), (c), and (e): The terms in the equation of motion for $\phi$~(\ref{equation_phi}) for different ranges of $r$. (b), (d), and (f): The corresponding evolutions of $\phi$. A strong $\psi$ can make the evolution of $\phi$ slow down, make a turn, and approach to $-\infty$.}
  \label{fig:collapse_physical_field}
\end{figure*}

So far, for simplicity, the parameter $\alpha$ has been set to $1$. As mentioned in Sec.~\ref{sec:introduction}, cosmological observations constrain $\alpha$ to be about $10^{-9}$, so next we consider vacuum collapse where $\alpha$ takes smaller values. It turns out that in this case more care is needed in setting up the initial conditions. As shown in Eq.~(\ref{potential_inflation_model}), for a fixed $\phi$, a small $\alpha$ will yield a large potential $V$. However, as implied in Eq.~(\ref{m_ic_collapse}), a large $V$ can make $m$ grow very fast with respect to $r$. As a result, $2m/r$ can approach $1$ at certain $r$, making the initial spacetime irregular. To avoid such an irregularity, we reduce the value of $\phi$ at $t=0$. To ensure that the system still has enough energy in the beginning so that collapse can happen in the future, we increase the spatial derivative of $\phi$ at $t=0$. We set $\alpha=0.01$. The initial profile for $\phi$ is defined as $\phi(x,t)|_{t=0}=a\cdot\cos(bx)$ for $bx\in[2\pi;~6\pi]$ and $0$ otherwise, with $a=0.04$ and $b=40$ (see Fig.~\ref{fig:evolutions_small_alpha}). In this configuration, a black hole can still be formed, and the radius of the apparent horizon is $r_{\scriptsize{\mbox{AH}}}\approx0.2$. As shown in Fig.~\ref{fig:evolutions_small_alpha}, in certain regions, $\phi$, $f'$ and $R$ can still be pushed to $+\infty$ as happens in the large $\alpha$ case. Using the same strategy and higher resolutions, we expect to be able to simulate collapse for $\alpha=10^{-9}$ and obtain similar results. The details are skipped here.

\section{Scalar collapse for the $R^2$ model\label{sec:scalar_collapse}}
Now we study scalar collapse. As in vacuum collapse, in scalar collapse, near the central singularity, the equations of motion for $\phi$~(\ref{equation_phi}) and the physical scalar field $\psi$~(\ref{equation_psi}) can be simplified as
\be \phi_{,tt}\approx-\frac{2}{r}r_{,t}\phi_{,t}-\frac{\kappa\psi_{,t}^2}{\sqrt{6}\chi}, \ee
\be \psi_{,tt}\approx-\frac{2}{r}r_{,t}\psi_{,t}+\sqrt{\frac{2}{3}}\kappa\phi_{,t}\psi_{,t}. \ee
Regarding the dynamics for $\phi$, the term $-2r_{,t}\phi_{,t}/r$ always enhances the growth of $|\phi_{,t}|$, while the term $-\kappa\psi_{,t}^2/(\sqrt{6}\chi)$ is always negative and tries to stop $\phi$ from growing. Regarding the dynamics for $\psi$, the term $\sqrt{2/3}\kappa\phi_{,t}\psi_{,t}$ enhances/slows down the growth of $|\psi_{,t}|$ when $\phi_{,t}$ is positive/negative.

We expect that, as long as $\psi$ is weak enough, the results of scalar collapse will be the same as those in vacuum collapse. Therefore, in this section, we skip the weak $\psi$ case and focus on the strong $\psi$ case. We find that when $\psi$ is strong enough and $\phi_{,t}$ is positive at certain moments, $|\psi_{,t}|$ can be accelerated to a large value. Consequently, $-\kappa\psi_{,t}^2/(\sqrt{6}\chi)$ can make $\phi$ slow down, make a turn, and approach to $-\infty$. In short, a strong physical scalar field can prevent $\phi$ and $R$ from growing to $+\infty$. This is shown in Fig.~\ref{fig:collapse_physical_field}. In making Fig.~\ref{fig:collapse_physical_field}, we have used the following configuration: $\alpha=0.01$, $\phi(x,t)|_{t=0}=a\cdot\exp[-(x-x_1)^2]$ with $a=-0.02$ and $x_1=3$, and $\psi(x,t)|_{t=0}=b\cdot\tanh(x-x_2)$ with $b=0.2$ and $x_2=3$.

\begin{figure*}
  \epsfig{file=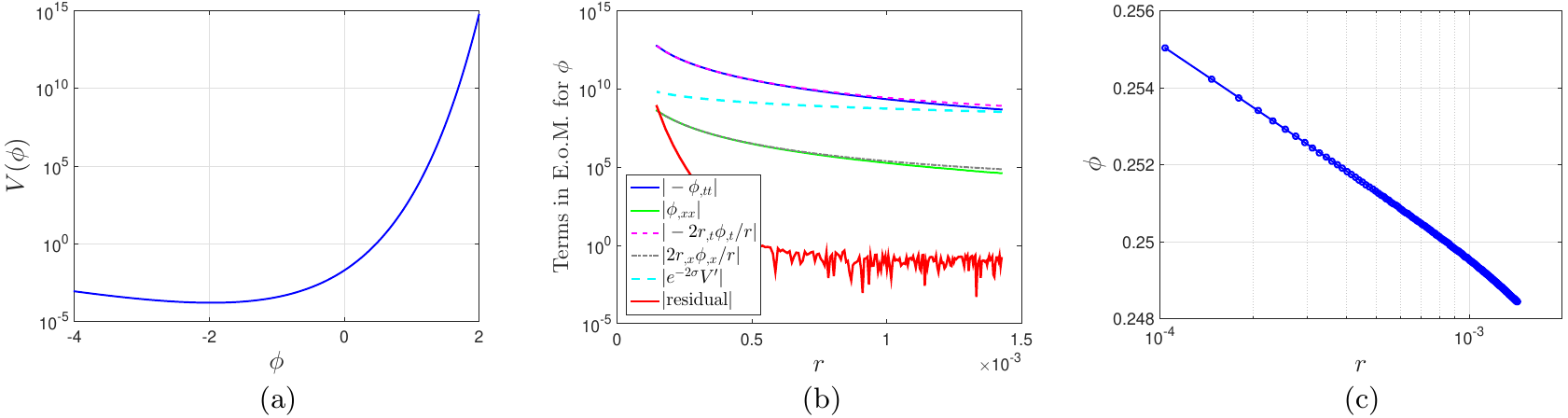, width=\textwidth}
  \caption{(color online). Vacuum collapse for the $R\ln{R}$ model. (a) Potential in the Einstein frame. (b) The terms in the equation of motion for $\phi$~(\ref{equation_phi}) for the slice $x=0.35$. (c) Evolution of $\phi$ for the slice $x=0.35$.}
  \label{fig:RlnR}
\end{figure*}

\section{Vacuum collapse for the $R\ln{R}$ model\label{sec:collapse_RlnR}}
It is meaningful to explore vacuum collapse for other $f(R)$ models so as to check whether it is universal that $R$ goes to infinity as the central singularity is approached. We take the $R\ln{R}$ model as one more example~\cite{Frolov_1101,Guo_1305}:
\be f(R)=R\left[1+\alpha\ln\left(\frac{R}{R_0}\right)\right],\ee
where $\alpha$ and $R_0$ are parameters. We set $\alpha=0.1$ and $R_{0}=1$. The potential for this model in the Einstein frame is plotted in Fig.~\ref{fig:RlnR}(a), which shows that the potential is very steep for large $\phi$. The initial profile for $\phi$ is set as $\phi(x,t)|_{t=0}=a\cdot\exp[-(x-x_0)^2/b]+c$, with $a=-0.1$, $b=c=0.02$, and $x_0=1$.

Vacuum collapse in the above configuration is simulated and the results are similar to those for the $R^2$ model. A black hole is formed, and in certain ranges of $x$, in the vicinity of the singularity, $f'$ goes to zero. As shown in Fig.~\ref{fig:RlnR}, in certain other ranges of $x$, near the central singularity, although the potential is very steep for large $\phi$, $\phi$, $f'$, and $R$ can still be pushed to infinity.

Now a natural question arises: Does an $f(R)$ model exist such that $\phi$ cannot be pushed to infinity? Our guess is no.

\section{Summary\label{sec:summary}}
We explored spherical vacuum and scalar collapse for the Starobinsky $R^2$ model. This model was built based on quantum corrections to the vacuum Einstein equations and is consistent with the Planck data on the cosmic microwave background. These quantum effects may be expected to avoid the singularity problem at the center of black holes and in the very early Universe. Moreover, the potential in the Jordan frame, where $f(R)$ gravity is originally defined, is very steep at high curvature scale. Thus one may expect that in collapse the steepness of the potential may prevent the Ricci scalar from approaching infinity. Our results show that in spherical collapse, when the scalar degree of freedom $f'$ or the matter field is strong enough, a black hole can be formed, and the singularity problem remains. These results are consistent with those obtained in Ref.~\cite{Hwang:2011kg}. In addition, we find that the steepness of the potential at high curvature scale in the Jordan frame is misleading to some extent. The potential in the Einstein frame is actually quite flat at high curvature scale. Due to strong gravity, in certain circumstances, $f'$ and the Ricci scalar can be pushed to infinity. In other words, the semiclassical approach such as the one used here seems not to avoid the singularity problem. In scalar collapse, a strong physical scalar field can prevent $f'$ from growing to infinity.

We found that in addition to scalar collapse, in vacuum collapse for the $R^2$ model, a black hole can also be formed, and the Ricci scalar can also be pushed to infinity as the central singularity is approached. Vacuum collapse for the $R\ln{R}$ model was also simulated. Although the potential for this model in the Einstein frame is steep at high curvature scale, the Ricci scalar is still pushed to infinity as the central singularity is approached. Consequently, such a feature seems universal in vacuum and scalar collapse in $f(R)$ gravity.

We note that the conclusions here have useful cosmological implications as well. This is because a time reversed case of the above investigation will correspond to an expanding universe model. Due to the similarity between the singularities in black holes and in the very early Universe, the quantum effects may not avoid the singularity problem in the very early Universe either, such as the $R^2$ model discussed here. Indeed, these arguments are in agreement with the results, obtained via perturbation analysis, on the existence of cosmological singularities for the $R^2$ model reported in Refs.~\cite{Ruzmaikina_1970,Barrow_1983}: Cosmological singularities cannot be avoided when the parameter $\alpha$ in the $R^2$ model (\ref{R_squared_model}) is positive.

\section*{Acknowledgments}%\small
The authors are very grateful to John D. Barrow, Sayantan Choudhury, Andrei V. Frolov, Ken-ichi Nakao, Tejinder P. Singh, and Dong-han Yeom for many helpful discussions. The authors appreciate the referees for their valuable comments. J.Q.G. thanks Simon Fraser University where part of this work was done.

%%%%%%%%%%%%%%%%%%%%%%%%%%%%%%%%%%%%%%%%%%%%%%%%%%%%%%%%%%%%%%%%

\end{document}